\documentclass[a4paper]{article}
\usepackage{Odyssey2022}
\usepackage{epsfig,amssymb,amsmath}
\usepackage{amsmath,graphicx}
\usepackage{multirow}
\usepackage{booktabs}
\usepackage{float}
\usepackage{subfigure}

\title{Low-Latency Online Speaker Diarization with Graph-Based Label Generation}

\makeatletter
\def\name#1{\gdef\@name{#1\\}}
\makeatother
\name{{\em Yucong Zhang$^1$, Qinjian Lin$^2$, Weiqing Wang$^{1}$, Lin Yang$^2$, Xuyang Wang$^2$, Junjie Wang$^2$, Ming Li$^{1,3}$}}

\address{$^1$Department of Electrial \& Computer Engineering, Duke University, Durham, NC 27708, USA\\
        $^2$AI Lab, Lenovo Research, Beijing 100085, China \\
        $^3$Data Science Research Center, Duke Kunshan University, Kunshan 215316, PR China \\
{\small \tt \{yucong.zhang, ming.li369\}@duke.edu} }
\begin{document}
\maketitle

\begin{abstract}
This paper introduces an online speaker diarization system that can handle long-time audio with low latency. We enable Agglomerative Hierarchy Clustering~(AHC) to work in an online fashion by introducing a label matching algorithm. This algorithm solves the inconsistency between output labels and hidden labels that are generated each turn. To ensure the low latency in the online setting, we introduce a variant of AHC, namely chkpt-AHC, to cluster the speakers. In addition, we propose a speaker embedding graph to exploit a graph-based re-clustering method, further improving the performance. In the experiment, we evaluate our systems on both DIHARD3 and VoxConverse datasets. The experimental results show that our proposed online systems have better performance than our baseline online system and have comparable performance to our offline systems. We find out that the framework combining the chkpt-AHC method and the label matching algorithm works well in the online setting. Moreover, the chkpt-AHC method greatly reduces the time cost, while the graph-based re-clustering method helps improve the performance. 
\end{abstract}

\section{Introduction}
\label{sec:intro}

Speaker diarization aims at solving the problem of "who spoke when". It is a process of separating the input audio into different pieces in terms of different speaker identities. Speaker diarization involving multiple speakers has various kinds of applications. Particularly, the boundaries produced by a diarization system can provide useful information to multi-speaker automatic speech recognition~\cite{kanda2019guided, medennikov2020stc} and improve its performance. 

The conventional modularized speaker diarization systems usually contain multiple modules~\cite{microsoft2021, landini2021analysis, multiscale2021}, including voice activity detection (VAD)~\cite{ng12b_interspeech}, speech segmentation, embedding extraction~\cite{dehak2010front, snyder2018x, wan2018generalized} and speaker clustering~\cite{garcia2017speaker, sell2014speaker, wang2018speaker, lin2019lstm}. 
Each of those modules has been studied widely to improve the overall performance of the diarization system. 
For embedding extraction, i-vector~\cite{dehak2010front}, x-vector~\cite{snyder2018x} and d-vector~\cite{wan2018generalized} are the most frequently used methods. 
In the clustering stage, embeddings are grouped together according to some similarity metrics. 
The typical clustering methods for speaker diarization are agglomerative hierarchy clustering (AHC)~\cite{garcia2017speaker, sell2014speaker} and spectral clustering (SC)~\cite{wang2018speaker, lin2019lstm}. 
In addition to the key modules mentioned above, re-clustering module may also be employed as the post-processing to further improve the performance~\cite{microsoft2021, landini2021analysis}.



\begin{figure*}[htbp]
	\centering
    \centerline{ \includegraphics[width=0.9\textwidth]{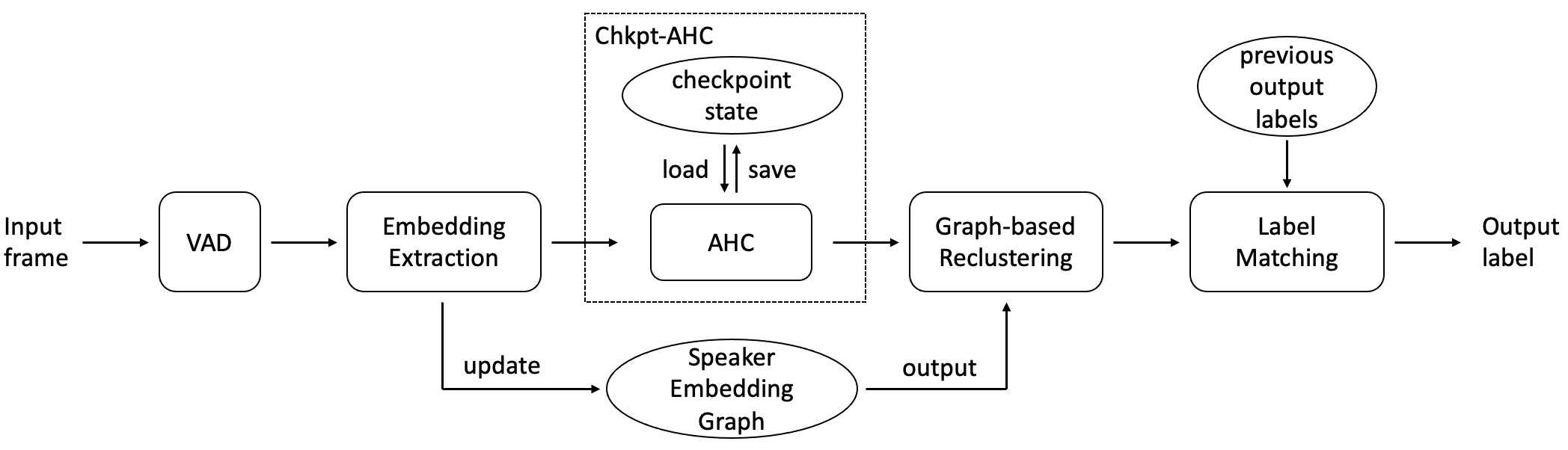} }
    \caption{The pipeline of our proposed system}
    \label{fig:pipeline}
\end{figure*}

Recently, the demand for online speaker diarization systems has surged in many scenarios, e.g., meeting or interview, but the conventional modularized speaker diarization system cannot be directly applied to the online diarization task, since most of the clustering algorithms are designed for offline tasks. An intuitive solution is that the clustering is performed on the whole received speech segments when new speech segments arrive. However, it is not time efficient and may cause high latency. Furthermore, the labels generated by the clustering algorithms might not be temporally consistent among all the speech segments. To handle all these problems, low-latency online diarization methods are studied.

Over the decade, several online speaker diarization systems have been developed. In the early design of the system, Gaussian Mixture Model (GMM) was trained as a background model, and some adaptation methods were used when a speech segment was assigned to a new speaker~\cite{geiger2010gmm, markov2008improved}. However, those systems usually need pre-trained models, such as GMM for male speech, female speech and non-speech. Later on, speaker embedding methods were proposed to replace the GMM approach. Some studies use d-vector as speaker embedding to represent speaker segments, and the embeddings were then clustered by some supervised methods~\cite{UISRNN2019, fini2020supervised}. However, those supervised diarization method needs lots of annotated diarization data, which 
might be difficult to acquire in real application. 

Besides supervised diarization methods, online modularized speaker diarization systems that use adapted i-vector and adaptive clustering are proposed~\cite{zhu2016online, dimitriadis2017developing}. Zhu et al.~\cite{zhu2016online} used principal component analysis (PCA) to transform speaker embeddings into a subspace, where the embeddings are more distinguishable. Dimitriadis et al. proposed a variation of the k-means algorithm several years ago~\cite{dimitriadis2017developing}. They refined the assignments of clusters by repeatedly attempting cluster subdivision, while keeping the most meaningful splits given a stopping criterion. However, the time complexity of both algorithms are linear with the number of speaker segments. Hence, it is not time-efficient to handle long-time audio clips.

More recently, an online speaker diarization system based on End-to-End Neural Diarization~(EEND)~\cite{han2021bw} was proposed. They modified the Encoder-Decoder-Attractor (EDA) architecture of Horiguchi et al.~\cite{horiguchi2020end} by adding an incremental Transformer encoder module. However, systems based on EEND have two major problems. First, it is restricted by the number of speakers. The online system purposed by Han et al.~\cite{han2021bw} performs comparably well to the offline systems when only one to two speakers are involved, but it cannot deal with more speakers well. Second, the end-to-end online neural diarization system needs a large amount of in-domain data to train beforehand. However, in-domain supervised training data for diarization cannot be easily obtained. Although simulated data can be used for training, the model still need to be transformed to the specific domain by finetuning on the real dataset. Those problems motivate us to build a modularized online speaker diarization system, which can deal with more speakers and without any training data.

In this paper, we propose an online modularized speaker diarization system, which can handle long-time audio without annotated training data. The source code will be released soon and a demo can be found in the Google Drive\footnote{https://drive.google.com/file/d/1v84QFzEsR7XeuNSV-h4WK-Qu1QRO3eBN/view?usp=sharing}. Our system is performed in a frame-wise online fashion that mainly consists of five modules, namely voice activity detection~(VAD), speaker embedding extraction, embedding clustering, post-processing and label generation. The system is illustrated in Fig.~\ref{fig:pipeline}. We summarize the key contribution of our online system as follows: \\

\noindent\textbf{i. Speaker clustering with chkpt-AHC} \\
We propose an online speaker clustering module, namely checkpoint-AHC (chkpt-AHC), to allow the whole system to process long-time audio with low latency. It is worth noting that this module functions normally in the offline setting, but it might cause label inconsistency problem in the online setting. Hence, in order to solve this problem, we propose a label matching algorithm to re-arrange the labels in the back-end of the system. We will explain more details in Section~\ref{subsubsec:label_perm} about the label inconsistency problem. \\

\noindent\textbf{ii. Post-processing with graph-based re-clustering} \\
After the clusters are derived from chkpt-AHC, we further refine the clusters using a speaker embedding graph, which is maintained and updated as new speaker embeddings arrive. In Section~\ref{subsubsec:DER}, the experimental results show that this module can slightly improve the performance. \\

\noindent\textbf{iii. Label matching} \\
In order to fix the label inconsistency issue, we introduce a novel approach to enable the online label generation for long-time audio, such that the output labels can remain the consistency. \\

The rest of the paper is organized as follows. Section~\ref{sec:proposed diarization system} presents the details of our online diarization system. Section~\ref{sec:experiments} describes the settings of our experiments and shows the results. Conclusions are drawn in Section~\ref{sec:conclusions}.




\begin{figure*}[htbp]
	\centering
    \centerline{ \includegraphics[width=0.8\textwidth]{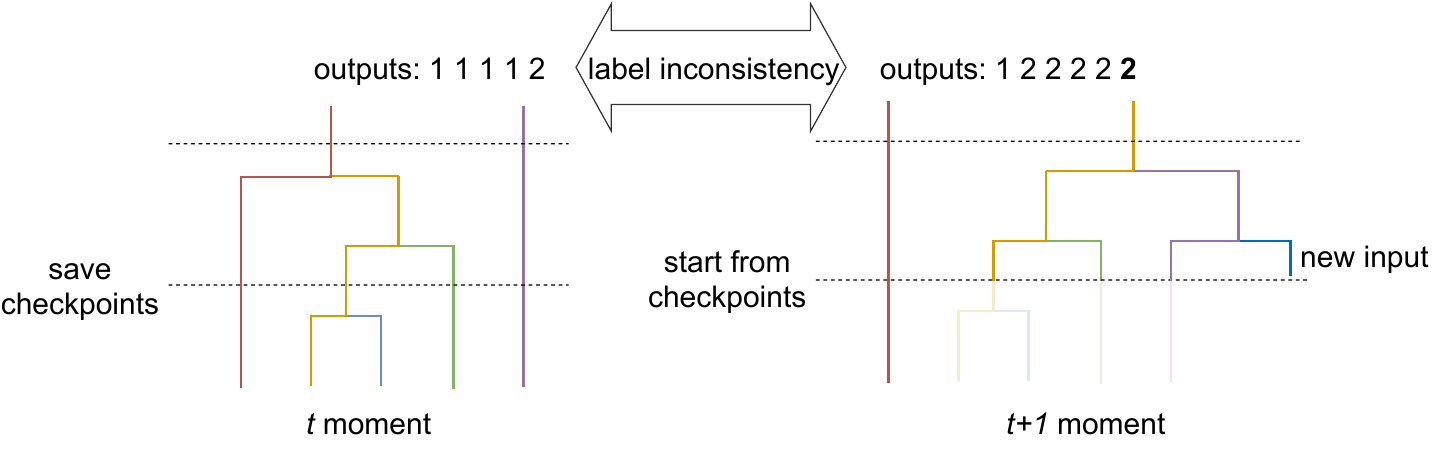}}
    \caption{Illustration of the label inconsistency problem}
    \label{fig:inconsistency}
\end{figure*}

\section{Proposed Diarization System}
\label{sec:proposed diarization system}

\subsection{Overview}
\label{subsec:overview}
The key idea of our work is to perform online clustering by only updating the label of the new speech segment based on the history segments.
The whole working pipeline is shown in Figure~\ref{fig:pipeline}. In the beginning, speech segments will be represented as speaker embeddings. Then, the embeddings will be clustered by a new variant of AHC algorithm, namely chkpt-AHC. In the meantime, a speaker embedding graph will be built and maintained accordingly. The resulting clusters generated by chkpt-AHC will be further processed by a graph-based re-clustering method. However, the labels generated after the clustering modules may not be consistent with previous output labels when a new speech segment appears, which is discussed in Section~\ref{subsubsec:label_perm}. Hence, we introduce a label matching module in the back-end, which adopts the Hungarian algorithm to align the speakers between two neighboring time steps. 

\begin{figure}[htbp]
	\centering
    \centerline{ \includegraphics[width=0.35\textwidth]{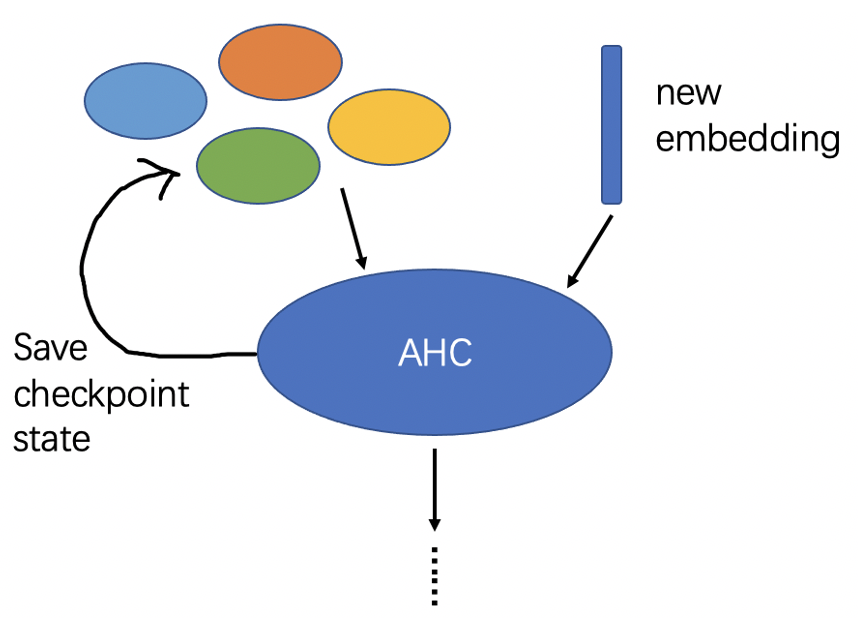} }
    \caption{chkpt-AHC}
    \label{fig:chkpt-AHC}
\end{figure}

\subsection{Online AHC and chkpt-AHC}
\label{subsec:chkpt-ahc}
The AHC algorithm for speaker clustering uses all the speaker embeddings, and it uses the centroid linkage criterion as a default setting, which means that each cluster is represented by its centroid embedding. Each time when a new speech segment arrives, the system first extracts its embedding and then performs AHC on all the previously saved speaker embeddings along with the new one. When AHC is performed, we use a pre-defined threshold $\theta$ as a stopping criterion, such that the similarity measures among all the resulting clusters are smaller than the threshold $\theta$. The original AHC algorithm is not friendly for online tasks, since the computational cost grows linearly with the length of the audio, leading to high latency for long audio clips.

Therefore, in order to reduce the computational cost, we set checkpoints to limit the number of initial speaker embeddings for the AHC algorithm, which is called chkpt-AHC.
When the number of speaker embeddings is fewer than the pre-defined checkpoint number $k$, the online system performs AHC on all of the speaker embeddings. Otherwise, the intermediate clustering results of $k$ clusters are recorded as a checkpoint state, and it will be used as the initial clustering state in the next time step. When the next speaker embedding arrives, the clustering process starts from the checkpoint state of $k$ clusters and continues the clustering process with the new-coming embedding. In this way, the checkpoint state is used to control the maximum number of speaker embeddings to be considered by AHC. With chkpt-AHC employed, the total processing time reduces significantly, especially for long-time audio, which is shown in Table~\ref{tab:time}.

\subsubsection{Label Inconsistency Problem}
\label{subsubsec:label_perm}
Although AHC-based clustering methods can be easily implemented and largely reduce the time for clustering, they cannot be solely applied to an online system. Simply adopting those methods will result in the label inconsistency problem. 

For simplicity, we call the labels that are generated right after the clustering modules as hidden labels, and the labels that are output by our system as output labels. 
Then, the label inconsistency problem refers to the inconsistency between output labels and hidden labels. 
The reason is that output labels cannot be changed once they are output, while hidden labels might change according to different inputs. 
Figure~\ref{fig:inconsistency} illustrates the label inconsistency problem. It shows the agglomeration process of chkpt-AHC with checkpoint number $k=4$. 
At $t^{th}$ moment, we have five speaker embeddings to agglomerate. 
Since the checkpoint number $k=4$, checkpoints are saved when the system agglomerates to four clusters. 
In the end, we can derive the speaker labels like ``11112", where each label stands for the index of the speaker. 
At $(t + 1)^{th}$ moment, as Section~\ref{subsec:chkpt-ahc} has mentioned, previous checkpoints will be loaded and clustered with the new input. 
Due to the change of the agglomerating order, the labels for the previous segments have changed to ``12222". 
However, ``11112" has already been output as output labels at $t^{th}$ moment, which are not able to change. 
In this case, the label inconsistency problem arises between two neighbor time steps. 
We are only allowed to find the best label for the new input embedding according to the current labels. 

To this end, we apply a label matching algorithm to solve this problem, which is described in Section~\ref{subsec:label matching}.


\subsection{Graph-based Re-clustering}
\label{subsec:graph-based}

Motivated by~\cite{microsoft2021}, we use a high threshold as the stopping criteria of chkpt-AHC to get high-purity clusters. As a result, the number of clusters after chkpt-AHC is usually larger than the ground-truth number of speakers. Then, we choose the speaker clusters based on the duration criterion. However, in the online setting, the duration of each cluster is always small at the beginning, since the duration of each cluster accumulates with time. Thus, we make some modifications so that the system can perform in an online manner. At first, when no cluster has a duration longer than a pre-defined threshold, we pick one cluster with the longest duration as the speaker cluster, and all the others as non-speaker clusters. Then, as the timestamp increases, the speaker clusters are picked using the duration criterion. In order to further refine the clustering results, we propose a graph-based approach to further determine whether the embeddings in non-speaker clusters belong to a current speaker cluster or a new speaker cluster.

\begin{figure}[hbp]
\begin{minipage}[b]{.48\linewidth}
  \centering
  \centerline{\includegraphics[width=4.0cm]{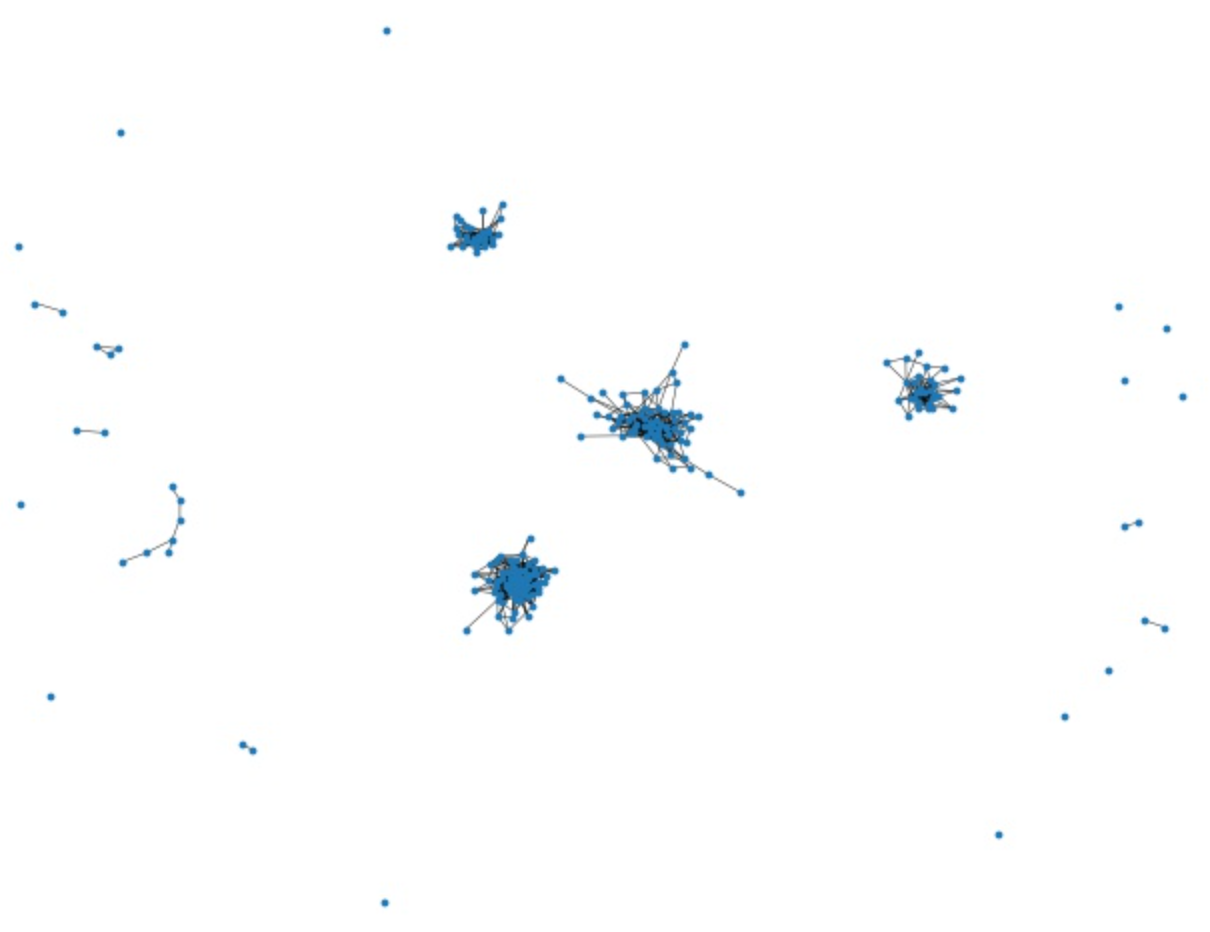}}
  \centerline{(a) Using 0.6 as threshold}\medskip
\end{minipage}
\hfill
\begin{minipage}[b]{0.48\linewidth}
  \centering
  \centerline{\includegraphics[width=4.0cm]{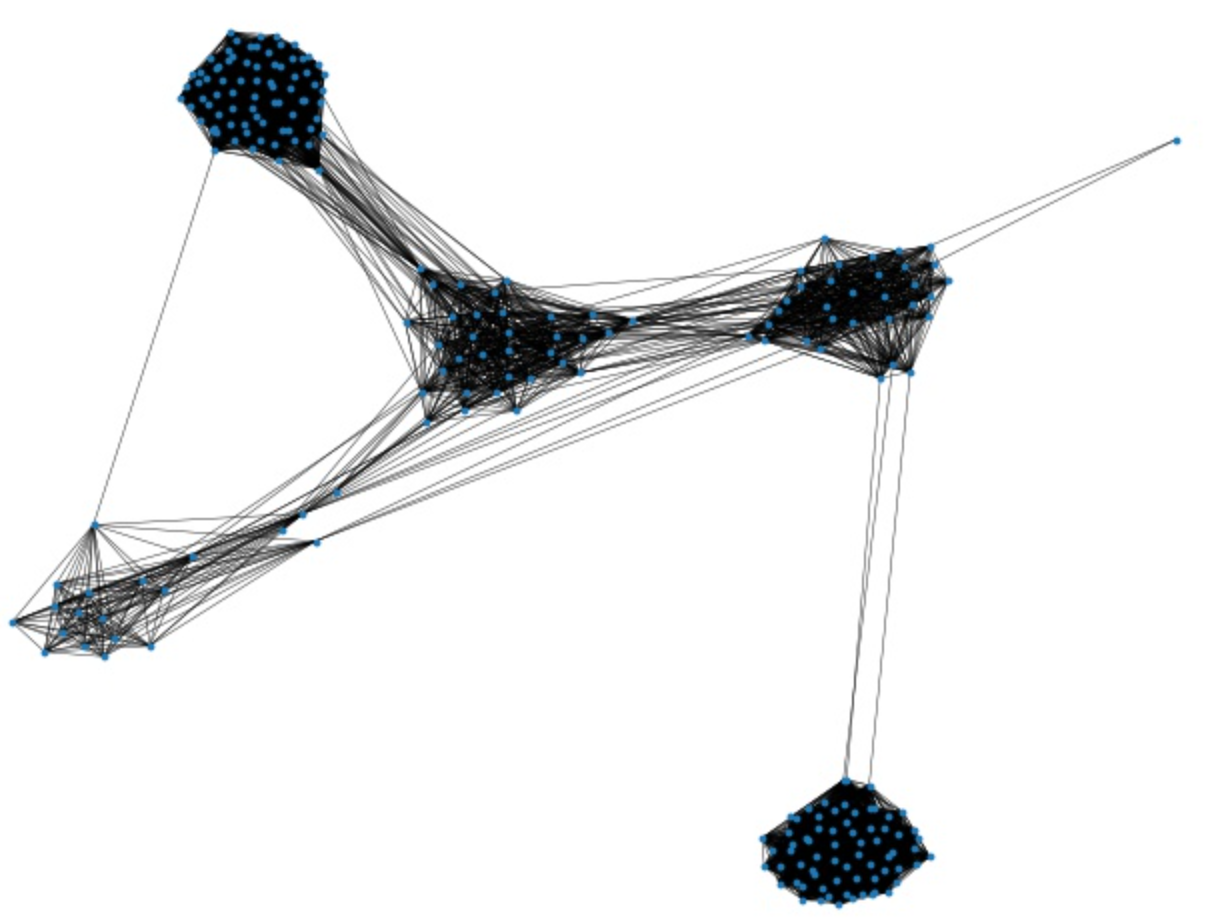}}
  \centerline{(b) Using 0.3 as threshold}\medskip
\end{minipage}
\caption{Graphs with different threshold}
\label{fig:graph thresholds}
\end{figure}

We introduce a speaker embedding graph $\mathcal{G}$, which is constructed as new speech segments arrive. Each node $n_i$ in the graph represents a speaker embedding $e_i$. The weight of the edges is the similarity between the speaker embeddings, measured by certain metric. The similarity threshold used to build a graph is lower than the stopping criteria of chkpt-AHC. As Fig.~\ref{fig:graph thresholds} shows, the graph contains more connected nodes when applying a lower threshold, which allows us to perform more precise refinements. Otherwise, it is difficult to adjust since many unconnected notes are present due to a high threshold. 

\begin{figure}[htbp]
\begin{minipage}[b]{.48\linewidth}
  \centering
  \centerline{\includegraphics[width=4.0cm]{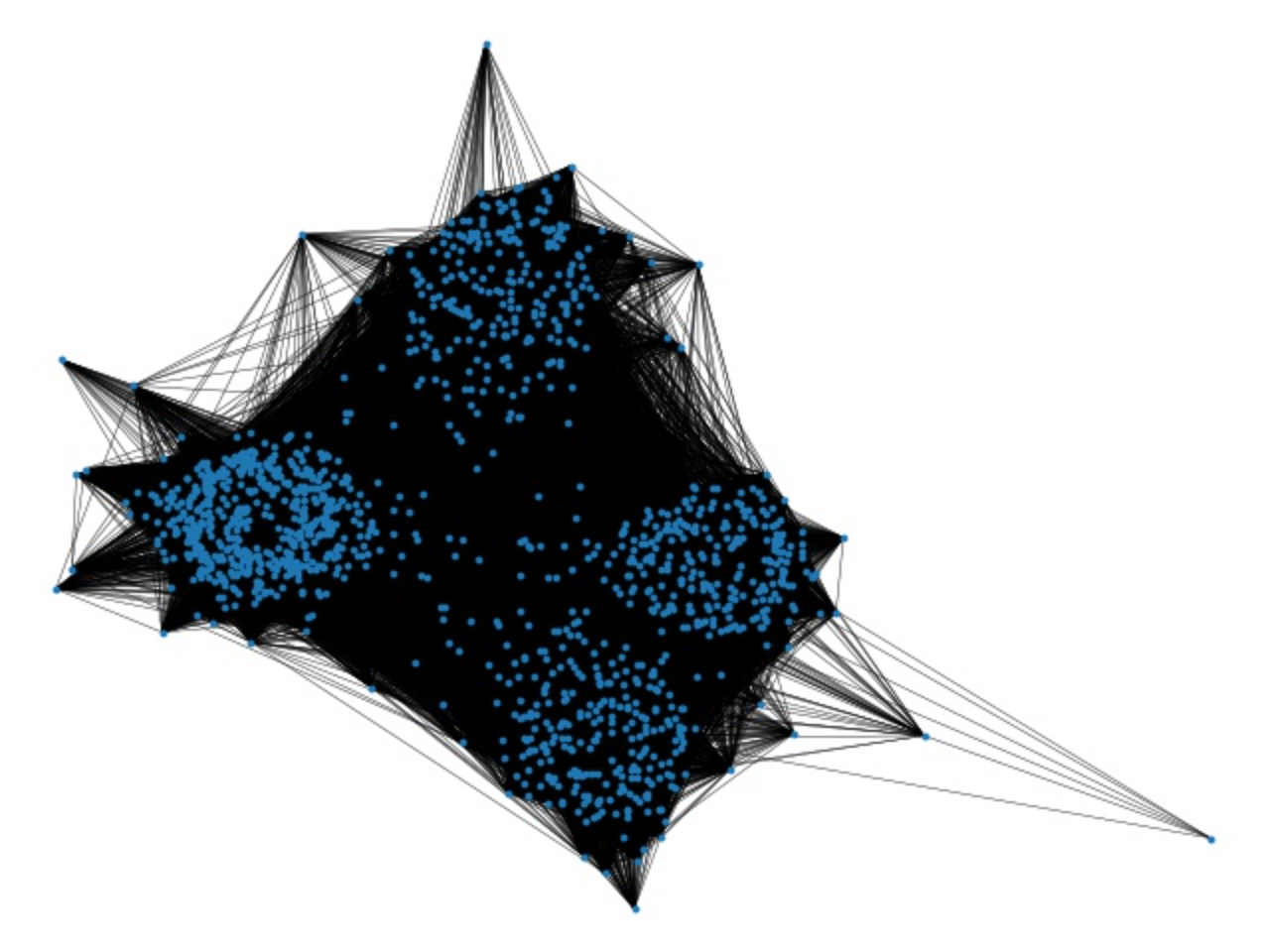}}
  \centerline{(a) Graph before pruning}\medskip
\end{minipage}
\hfill
\begin{minipage}[b]{0.48\linewidth}
  \centering
  \centerline{\includegraphics[width=4.0cm]{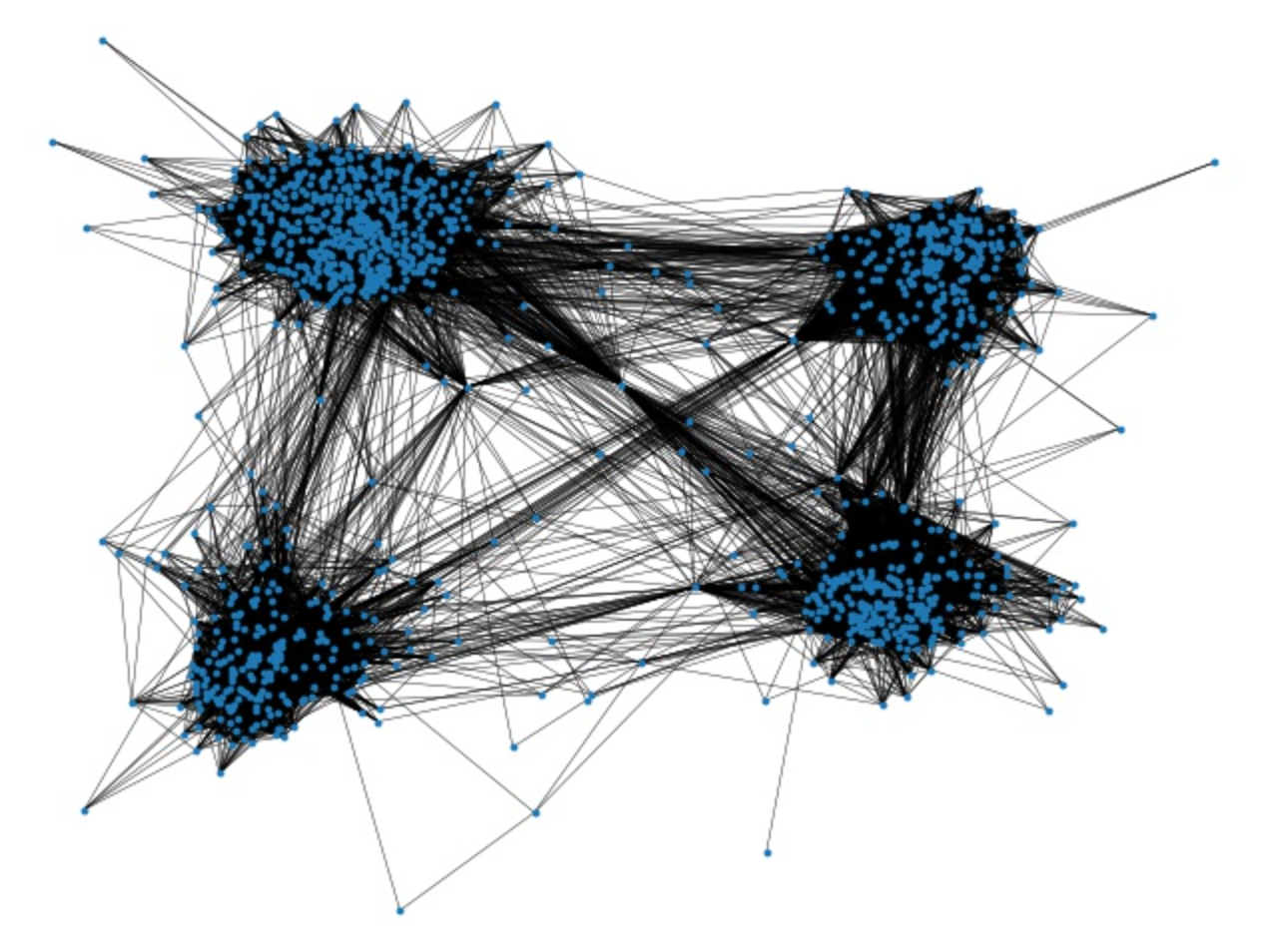}}
  \centerline{(b) Graph after pruning}\medskip
\end{minipage}
\caption{Effects of Graph pruning with threshold=0.3}
\label{fig:graph pruning}
\end{figure}

Since building such a graph is computationally expensive, we prune some of the edges in the speaker embedding graph as shown in Fig.~\ref{fig:graph pruning}. Intuitively, if the new speaker node is not similar to another node in the graph, the new node is not connected to that node and that node's neighbors in the graph.

With the auxiliary speaker embedding graph, we propose a novel approach to deal with the embeddings in non-speaker clusters. Given a speaker embedding graph $\mathcal{G}$, we can represent it as a weighted adjacency matrix $A$ as follows:
$$
A_{ij} = \left\{\begin{array}{cc}
    \theta_{ij}, & \theta_{ij} > s \\
    0, & \theta_{ij} \le s
\end{array}\right.
$$
where $\theta_{ij}$ is the similarity score between embedding $e_i$ and embedding $e_j$. We use cosine similarity as the score metric in the experiment. \\
\noindent We define the cluster likelihood $\mathcal{L}_{j}^{(i)}$ to measure how likely node $n_i$ belongs to the $j^{th}$ speaker cluster. The $j^{th}$ speaker cluster is represented by a set $C_j$ that contains the indices of the nodes in the $j^{th}$ cluster. The cluster likelihood $\mathcal{L}_{j}^{(i)}$ can then be calculated as follows:
$$
\mathcal{L}_{j}^{(i)} = \frac{\sum_{k \in C_j} A_{ik}}{\left|C_j\right|}
$$
where $\left|C_j\right|$ is the cardinality of $C_j$, counting the number of nodes in the $j^{th}$ cluster.


With the speaker embedding graph and cluster likelihood, we can assign a non-speaker 
node to the cluster with the highest cluster likelihood:
$$
l(n_i) = \text{argmax}_{j}\mathcal{L}_{j}^{(i)}
$$
where $l(n_i)$ is the index of the cluster that node $n_i$ should be assigned to.

\subsection{Label Matching Algorithm}
\label{subsec:label matching}

\begin{figure}[hbpt]
\begin{minipage}[b]{.48\linewidth}
  \centering
  \centerline{\includegraphics[width=4.5cm]{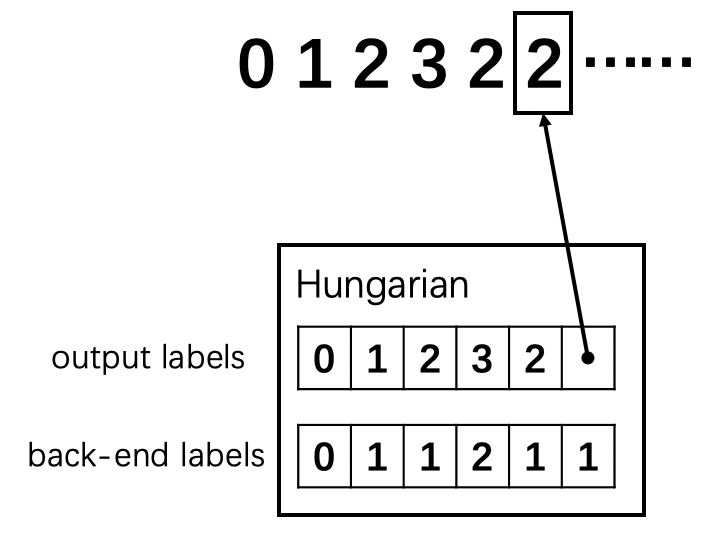}}
  \centerline{(a) Label matching}\medskip
\end{minipage}
\hfill
\begin{minipage}[b]{0.48\linewidth}
  \centering
  \centerline{\includegraphics[width=3.0cm]{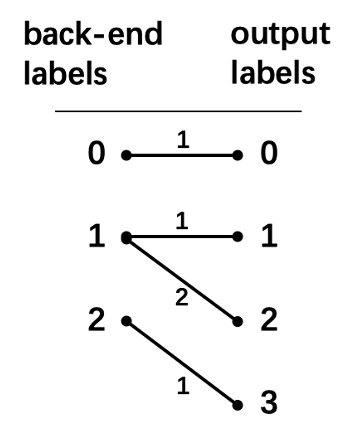}}
  \centerline{(b) Bipartite graph}\medskip
\end{minipage}
\caption{Front-end module}
\label{fig:label generating}
\end{figure}

Labels of each speech segment are generated according to different clusters in the back-end module. However, with the appearance of new speech segments, the clustering result might not be consistent, which is illustrated in Section~\ref{subsubsec:label_perm}. Hidden labels generated in the back-end might change when new speech segments arrive, but the previously output labels in the front-end are not allowed to change. This leads to a mismatch between output labels and hidden labels as shown in Fig.~\ref{fig:label generating}~(a). This motivates us to implement an algorithm to match the back-end labels and the output labels so as to make inferences on the label of the new speech segment. We re-frame the label matching problem as a maximum weighted bipartite matching problem, which can be solved using the Hungarian algorithm~\cite{kuhn1955hungarian}. Each node in the bipartite graph stands for a label. Output labels and back-end labels are in the separate part of the graph. The edge weight between two nodes is the frequency that two labels appear at the same time. An example of the graph is shown in Fig.~\ref{fig:label generating}~(b). 

By formulating the problem as a bipartite matching problem, we derive the matching result using the Hungarian algorithm. We then use this result to make inferences on the new speech segment as shown in Fig.~\ref{fig:label generating}~(a).

\section{Experiments}
\label{sec:experiments}
We first build three baseline speaker diarization systems, including two offline systems and one online system. All three baseline systems are described in detail in Section~\ref{subsec:baseline_sys}. Then, we measure the performance of our proposed online system and compare it with the performance of all baseline systems. Finally, we record the processing time of our proposed online systems in order to measure the time efficiency.

Throughout the whole experiment, offline speaker diarization systems are only tested in the offline setting, while online ones are only tested in the online setting.

\subsection{Datasets}
\label{subsec:dataset}
For the speaker embedding model, it is trained on the development set of Voxceleb2~\cite{nagrani2017voxceleb} with 5994 speakers and achieves an equal error rate (EER) of 1.06\% on the Voxceleb1 original test set. For the speaker diarization part, we use full sets in DIHARD3~\cite{ryant2020third} challenge and VoxConverse~\cite{chung2020spot} datasets as our diarization datasets. For DIHARD3, there are 254 audio clips in the development dataset and 259 audio clips in the evaluation dataset. For VoxConverse, there are 216 audio clips in the development dataset and 232 audio clips in the evaluation dataset. We use the development datasets to tune the parameters and evaluate our systems on the evaluation datasets. 

{\ninept
\begin{table*}[htbp]
\centering
\caption{The DER (\%) of the proposed speaker diarization system. }
\vspace{2mm}
\begin{tabular}{c|cc|ccccc|cc|cc}
\toprule
\multirow{2}{*}{\textbf{Sys.}} & \multirow{2}{*}{\textbf{Offline}} & \multirow{2}{*}{\textbf{Online}} & \multirow{2}{*}{\textbf{AHC}} & \multirow{2}{*}{\begin{tabular}[c]{@{}c@{}}\textbf{Chkpt-}\\ \textbf{AHC}\end{tabular}} & \multirow{2}{*}{\begin{tabular}[c]{@{}c@{}}\textbf{Naive}\\ \textbf{re-clustering}\end{tabular}} & \multirow{2}{*}{\begin{tabular}[c]{@{}c@{}}\textbf{Graph-based}\\ \textbf{re-clustering}\end{tabular}} & \multirow{2}{*}{\begin{tabular}[c]{@{}c@{}}\textbf{
Label}\\ \textbf{Matching}\end{tabular}} & \multicolumn{2}{c|}{\textbf{DIHARD3}} & \multicolumn{2}{c}{\textbf{VoxConverse}} \\\cmidrule(lr){9-10}\cmidrule(lr){11-12}
 & & & & & & & & \textbf{Dev.} & \textbf{Eval.} & \textbf{Dev.} & \textbf{Eval.} \\ \midrule
Base.~1 & $\surd$ & - & - & - & - & - & - & 20.71 & 20.75 & - & - \\
Base.~2 & $\surd$ & - & $\surd$ & - & $\surd$ & - & - & 17.63 & 16.82 & 3.94 & 4.68 \\ 
Base.~3 & - & $\surd$ & - & - & - & - & - & 39.07 & 36.79 & 10.34 & 14.65 \\
\midrule
Prop.~1 & - & $\surd$ & $\surd$ & - & $\surd$ & - & $\surd$ & \textbf{20.17} & 19.68 & \textbf{5.20} & \textbf{6.28} \\
Prop.~2 & - & $\surd$ & - & $\surd$ & $\surd$ & - & $\surd$ & 20.78 & 20.05 & 5.91 & 6.71 \\
Prop.~3 & - & $\surd$ & - & $\surd$ & - & $\surd$ & $\surd$ & 20.28 & \textbf{19.57} & 5.80 & 6.60 \\
\bottomrule
\end{tabular}
\label{tab:DERs}
\end{table*}
}

\subsection{Baseline Systems}
\label{subsec:baseline_sys}
We include three baseline systems for comparison as shown in Table.~\ref{tab:DERs}. Two of them are offline systems and another one is an online system. Here is the detailed description: 

\noindent\textbf{Baseline~1} \\
The first baseline system is an offline speaker diarization system. It is introduced by~\cite{ryant2020third} for DIHARD3 competition without VB-HMM resegmentation. 

\noindent\textbf{Baseline~2} \\
The second baseline system is an offline speaker diarization system made on our own. It uses a similar AHC-based speaker clustering module with the naive re-clustering module as shown in~\cite{microsoft2021}. The naive re-clustering module uses a threshold to reassign the non-speaker clusters to the speaker clusters. Intuitively, the non-speaker clusters are assigned to one of the speaker clusters via cosine similarity between centroid embeddings with a threshold. Our implementation of the offline speaker diarization system has the DER of 16.82\% on the evaluation dataset of DIHARD3, which outperforms the official baseline provided in the DIHARD3 competition~\cite{ryant2020third}. Moreover, the result is better than half of the teams that take part in the DIHARD3 competition, which indicates that our offline system has a comparable performance with the state-of-the-art offline systems. 

\noindent\textbf{Baseline~3} \\
The third baseline system is an online system. It contains none of our proposed modules, and the mechanism behind it is intuitive. When a new speech segment arrives, this system will assign it to the most similar cluster and update the centroid embedding of that cluster. Here, we measure the similarity between the new speech segment and a cluster by calculating the cosine similarity between the speaker embedding of the new segment and the centroid embeddings of the clusters. Especially, if the similarity score is lower than a pre-defined value, the new speech segment will form a new cluster on its own.

\subsection{Model Configurations}
\label{subsec:model config}

\subsubsection{Speaker Embedding Extraction}
\label{subsubsec:embedding extraction}
In our work, we use the same recipe described in~\cite{cai2020fly}. The input frame is 1s in length with a 0.5s shift. We use a deep CNN based on ResNet~\cite{he2016deep} to extract the 128-dim speaker embeddings.

\subsubsection{Thresholds for Speaker Clustering and Re-clustering}
\label{subsubsection:speaker clustering}
\begin{figure}[htbp]
\begin{minipage}[b]{0.4\linewidth}
  \centering
  \centerline{\includegraphics[width=4.5cm]{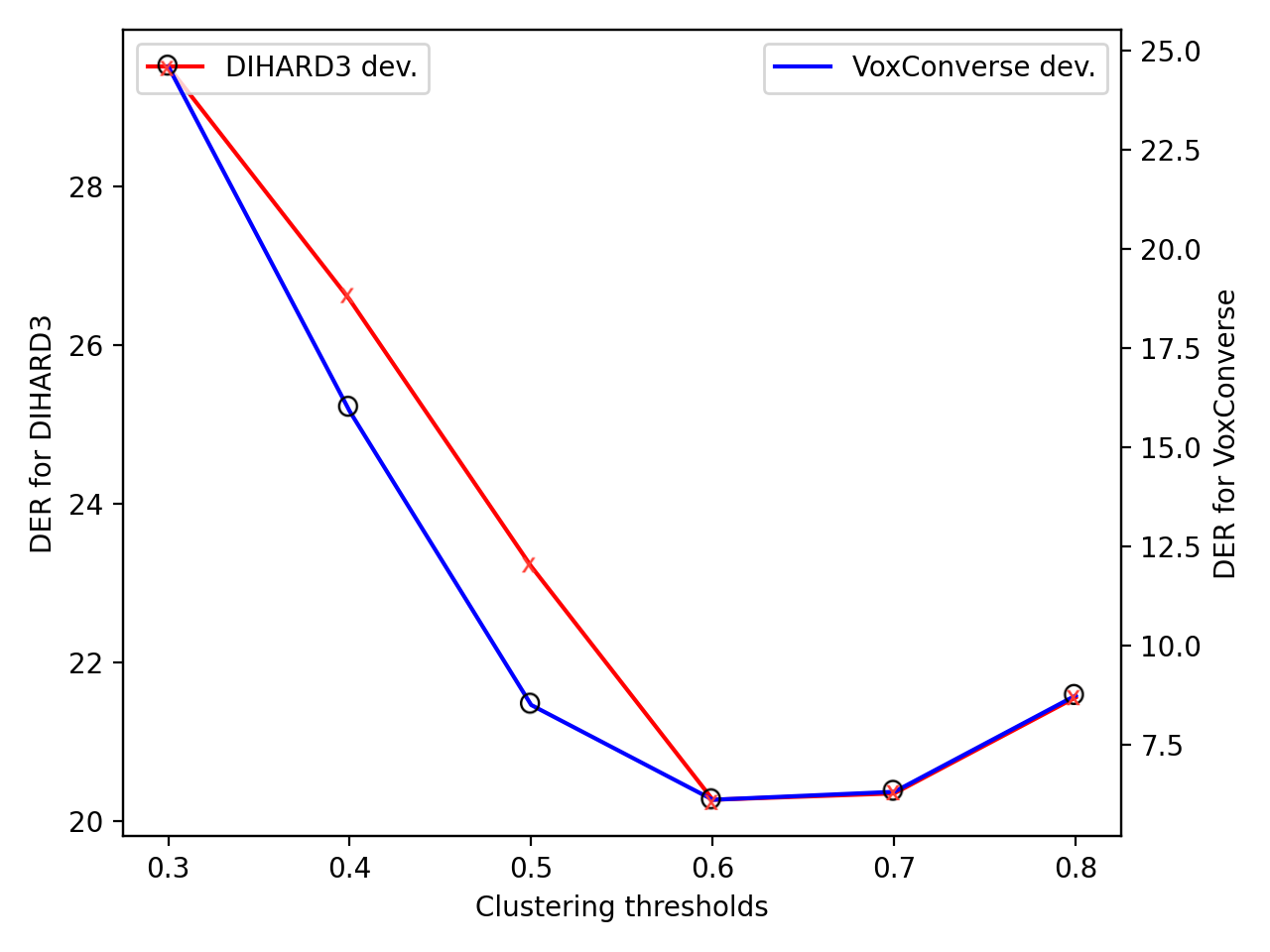}}
  \centerline{(a) chkpt-AHC}\medskip
\end{minipage}
\hfill
\begin{minipage}[b]{0.4\linewidth}
  \centering
  \centerline{\includegraphics[width=4.5cm]{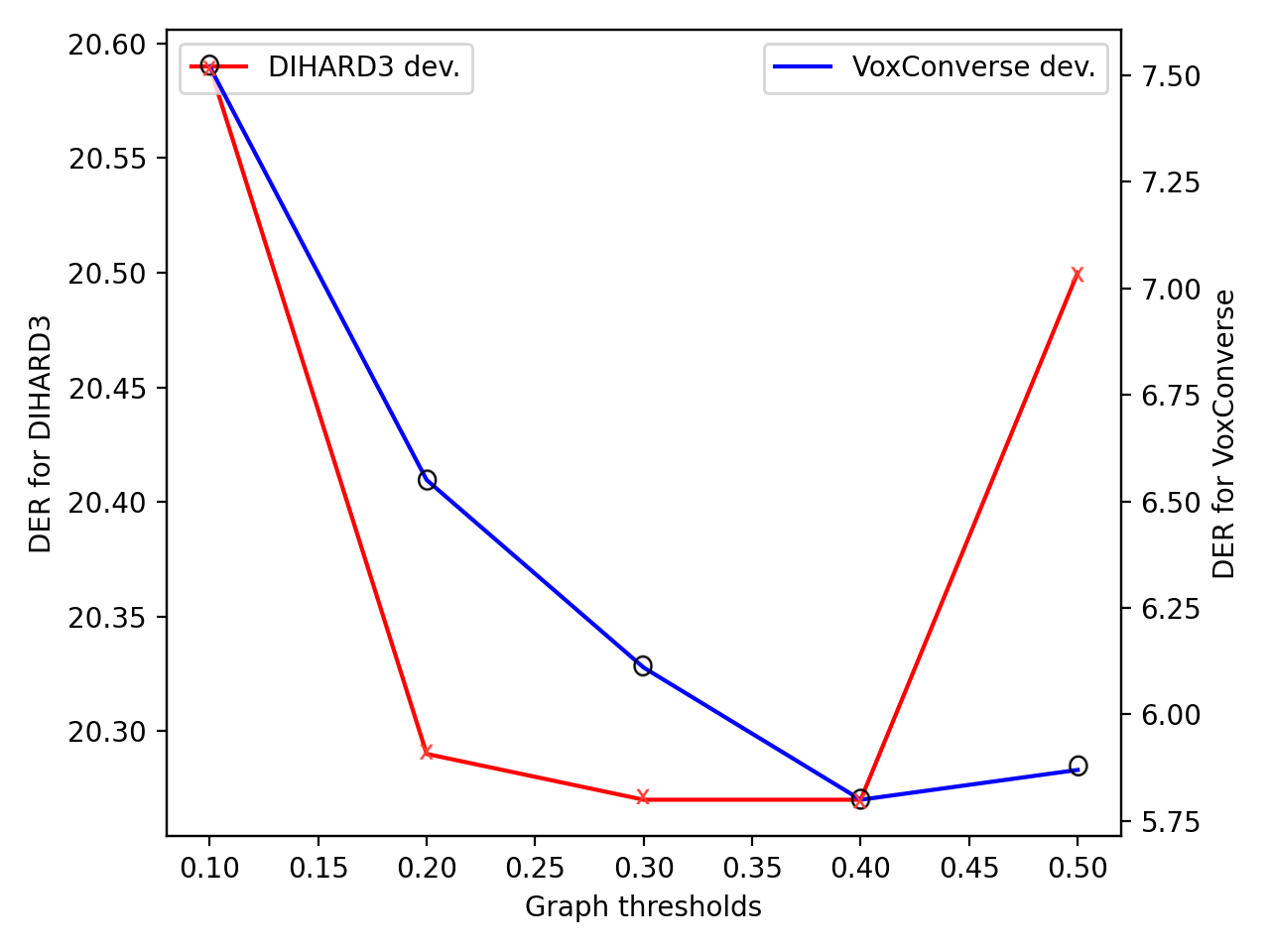}}
  \centerline{(b) speaker embedding graph}\medskip
\end{minipage}
\caption{Thresholds tuning}
\label{fig:thresholds}
\end{figure}

The thresholds for chkpt-AHC and speaker embedding graph are tuned on the corresponding development datasets, as shown in Fig.~\ref{fig:thresholds}. The optimal thresholds for different datasets are similar, which shows that similar thresholds can be applied to different datasets. As a result, we use 0.6 as the stopping threshold for both AHC and chkpt-AHC, and we use 0.4 as the threshold to build the speaker embedding graph for the experiment. In addition, we limit the maximum number of clusters for chkpt-AHC to 50 in our experiment.

\subsection{Evaluation Metrics}
\label{subsec:metrics}
We use the diarization error rate (DER) to measure the performance of our system. DER typically consists of three components: False Alarm (FA), Speaker Confusion and Missed Detection (Miss). Particularly, for the Voxconverse dataset, we evaluate our model with a 0.25s forgiveness collar for DER.

\subsection{Results \& Analysis}
\label{subsec:results}


\subsubsection{DER Performance}
\label{subsubsec:DER}

The overall diarization results are shown in Table~\ref{tab:DERs}. We do not provide the DER results on VoxConverse of baseline system~1, since oracle VAD is used. In our proposed online system~1, We build an online speaker diarization system that uses AHC and a naive re-clustering module with a label matching module in the end. This system has similar offline modules of the second baseline system. The key difference is that we use a label matching algorithm, which enables the state-of-the-art offline diarization components, such as the AHC and the re-clustering module, to be embedded in an online system. Table~\ref{tab:DERs} shows that by enabling the AHC and the re-clustering module, the performance of our proposed online system~1 is much better than the baseline online system. Moreover, the performance is even better than the offline baseline system~1.

In our proposed system~2, we further change the AHC module with the chkpt-AHC module and leave the re-clustering module unchanged. Compared to the AHC module used in our proposed system~1, the clustering accuracy drops in the sense that chkpt-AHC starts clustering from the intermediate checkpoint state instead of clustering from the beginning. It is reasonable to see that the DER goes up a little bit. Despite the tiny drop in the performance, our proposed system~2 is much more time-efficient than our proposed system~1 as illustrated in Section~\ref{subsubsec:time}.

Based on our proposed system~2, proposed system~3 replace the naive re-clustering module with a graph-based re-clustering module. Compared to proposed system~2, proposed system~3 gains improvement on all the datasets, which indicates that the graph-based re-clustering module works better than the naive re-clustering module. 

\noindent\textbf{(Online vs. Online)} 
We compare the performance of our proposed online systems with the baseline online system. As shown in Table~\ref{tab:DERs}, all our proposed online systems outperform the baseline online system~3 by a large margin. The results indicate that the framework including offline speaker clustering methods, such as AHC or chkpt-AHC, and the label matching algorithm works well in the online setting. 

\noindent\textbf{(Offline vs. Online)} 
Although our proposed online systems perform a little bit worse than the state-of-the-art baseline offline system, they still outperform the baseline offline system~1, which indicates that our proposed online systems can achieve comparable performance to the offline systems.

{\ninept
\label{subsec:result}
\begin{table*}[htbp]
\centering
\caption{Average processing time~(s) / average audio time~(s) for different datasets. This experiment is conducted on a single core of Intel(R) Xeon(R) CPU E5-2630 v4 @ 2.20GHz.}
\vspace{2mm}
\begin{tabular}{@{}c|cccc@{}}
\toprule
\multirow{2}{*}{\textbf{Sys.}} & \multicolumn{2}{c}{\textbf{DIHARD3}} & \multicolumn{2}{c}{\textbf{VoxConverse}} \\ \cmidrule(l){2-5}
 & \textbf{Dev.} & \multicolumn{1}{c|}{\textbf{Eval.}} & \textbf{Dev.} & \textbf{Eval.} \\ \midrule
Prop.~1 & 121.8/484.0 & \multicolumn{1}{c|}{113.0/458.8} & 104.3/338.3 & 446.1/830.2 \\
Prop.~2 & 44.0/484.0 & \multicolumn{1}{c|}{40.0/458.8} & 27.9/338.3 & 114.7/830.2 \\
Prop.~3 & 48.9/484.0 & \multicolumn{1}{c|}{44.7/458.8} & 32.5/338.3 & 142.7/830.2 \\ \bottomrule
\end{tabular}
\label{tab:time}
\end{table*}}

\subsubsection{Time Efficiency}
\label{subsubsec:time}

Online speaker diarization systems need to handle speech data in real-time scenarios and make lively updates, so besides the DER performance, time efficiency is also an important criterion. We use the average processing time that a system takes to handle one audio clip in the datasets as a metric to measure the time efficiency. Intuitively, it is calculated by the time of processing the whole dataset divided by the total number of audio clips in the dataset.

We compare the time efficiency among our proposed systems. Table.~\ref{tab:time} shows that the time cost for our proposed system~1 is considerably high compared to the other two, because it takes all the embeddings into account when clustering. Our proposed system~3 is a little bit more time-consuming than the proposed system~2, because we add a speaker embedding graph to the system. Building and maintaining such a graph might take extra time. Overall, we find out that by applying chkpt-AHC, the online system is much more time-efficient than the online system with original AHC, and the performance is almost the same. To further show that our proposed system~3 is time-efficient, a demo can be found in Google Drive\footnote{https://drive.google.com/file/d/1v84QFzEsR7XeuNSV-h4WK-Qu1QRO3eBN/view?usp=sharing}.




\section{Conclusions}
\label{sec:conclusions}
In this paper, we propose an online modularized diarization system that can handle long-time audio with low latency. In the system, we propose a label matching algorithm to handle the label inconsistency problem, which enables us to embed the offline speaker clustering approaches, such as AHC, in an online system. We further propose the chkpt-AHC method and the graph-based re-clustering method. Through the experiments, we find out that applying chkpt-AHC significantly improves time efficiency, and adopting the graph-based re-clustering method helps improve the performance. We experimentally show that our proposed online systems achieve better performance than our baseline online system, and comparable performance with the offline systems.

\vfill\pagebreak

\bibliographystyle{IEEEbib}
\bibliography{main}

%

\end{document}